\documentclass[%
amsmath,
reprint,
prc,
]{revtex4-1}

\usepackage[T1]{fontenc}
\usepackage{textcomp}
\usepackage{txfonts}
\usepackage{bm}
\usepackage{graphicx}
\usepackage{dcolumn}
\usepackage{xcolor}
\usepackage{lipsum} 
\usepackage{mathtools}
\usepackage{sidecap}
\usepackage{booktabs}
\usepackage{float}
\usepackage{ragged2e}
\usepackage{anyfontsize}
\usepackage{braket}
\usepackage{hyperref}

\usepackage{chngcntr}

\hypersetup{colorlinks=true, linkcolor=blue, citecolor=blue, urlcolor=blue}

\begin{document}

\title{Impact of shape coexistence on the symmetric to asymmetric fission mode transition in Th isotopes}

\author{Shengyuan Chen$^{1}$}
\author{Zeyu Li$^{2,1}$}
\author{MingHhui Zhou$^{1}$}
\author{Zhipan Li$^{1}$}
\email{E-mail: zpliphy@swu.edu.cn}
\affiliation{%
$^{1}$School of Physical Science and Technology, Southwest University, Chongqing 400715, China\\
$^{2}$China Nuclear Data Center, China Institute of Atomic Energy, Beijing 102413, China
}

\date{\today}

\begin{abstract}
\fontsize{10pt}{12.6pt}\selectfont
We study the evolution of fission modes along the Th isotopic chain using a microscopic framework combining the time-dependent generator coordinate method and finite-temperature covariant density functional theory. Theoretical fission fragment charge distributions agree well with experiments, and reveal a rapid symmetric-to-asymmetric transition from $A=222$ to 234. By analyzing the collective potential energy surfaces and time evolution of collective probability density distributions, we demonstrate that this fission mode transition is strongly correlated with the rapidly deepening asymmetric fission valley --- a phenomenon driven by the reduction of deformation energies of both the heavy and light fragments formed in the asymmetric fission valley. Further analysis attributes the decrease of light-fragment deformation energies to the onset of a coexisting large-deformed minimum in neutron-rich Kr and Sr isotopes (the dominated isotopes for light asymmetric peak), which arises from a deformed proton $Z=38$ shell closure near $\beta_2\approx0.46$. Notably, we identify, for the first time, the pivotal role of the light fragment and its shape coexistence structure on the fission mode transition in Th isotopes in a fully microscopic framework. 
\end{abstract}

\maketitle

\fontsize{11pt}{13.2pt}\selectfont
\section{INTRODUCTION}
Nuclear fission, in which nucleus splits into two or more fragments, has been thoroughly studied for more than 85 years since its discovery \cite{Hahn1939Jan,Meitner1939Mar,Bohr1939Sep}, but some key aspects of the fission dynamics are still incompletely understood. One notable feature is the rapid transition between symmetric and asymmetric fission modes as the mass number of the compound nucleus changes \cite{Andreyev2013Oct,Ryzhov2011May,Schmidt2000Feb,Hoffmann1992Oct,Bogachev2021Aug,Andreyev2018,Schmidt2018Sep}. Typical symmetric charge (mass) distributions are observed in the fission of pre-actinides, whereas asymmetric distributions are known to characterize the fission of U-Cf nuclei \cite{Andreyev2013Oct,Schmidt2000Feb}. Notably, certain isotopic chains (e.g., Th, Fm) display a rapid transition from symmetric to asymmetric fission modes \cite{Schmidt2000Feb,Chatillon2019May,Hoffmann1992Oct,Harbour1973Oct,Gindler1977Oct,Flynn1972May,Flynn1975Nov,Hulet1989Aug,Hoffman1980Mar}. For instance, in even-even Th isotopes, the fission fragment charge distribution (FFCD) evolves from a single symmetric peak in $^{222}$Th, to a triple-peak structure in $^{226}$Th, and finally to asymmetric peaks dominating in $^{228,230}$Th. In particular, a novel experimental setup has been developed that enables unambiguous identification of fissioning nuclei with mass number $A$, $A-1$, etc., thereby substantially suppressing the influence of higher-chance fission on the fission fragment yield distributions \cite{Chatillon2019May}. This provides an excellent platform for investigating the rapid transition between fission modes and underlying the microscopic mechanism.

Theoretically, the macroscopic-microscopic model \cite{Randrup2011Mar,Pomorski2021May,Randrup2013Dec,Nerlo2013May,Pashkevich2008Sep}, scission point model \cite{Pasca2016Dec,Pasca2016Sep,Ruan2019Dec}, dynamical cluster-decay model \cite{Jain2022Mar}, etc. have been applied to investigate the rapid evolution of the fission mode in Th isotopic chain. The main features of the symmetric/asymmetric transitions of the FFCD are basically reproduced, and it is suggested that the transition between the fission modes is influenced by the fragments formed in the asymmetric fission valley, which align closely with the proton/neutron numbers $N_H$ = 82 and $N_L, Z_H$ = 50 \cite{Pasca2016Sep}. However, a clear explanation of the underlying microscopic mechanism of the symmetric/asymmetric transition is notably absent in these methods and requires further study within microscopic theory.

Nuclear density functional theory (DFT) exhibit promising potential in both the qualitative and quantitative description of fission dynamics and fission data \cite{Schunck2016Nov,Schmidt2018Sep,Simenel2018Nov,Bender2020Nov,Verriere2020Jul,Schunck2022Jul}. The calculation and analysis of microscopic potential energy surfaces have identified the existence of symmetric and asymmetric fission paths, pointing out their competitive nature \cite{Berger2000Jul,Bernard2020Apr,Dubray2008Jan,Mcdonnell2013May}. Nevertheless, these calculations lack consideration of the dynamical effects, which are generally realized through two main methods: the time-dependent generator coordinate method (TDGCM) \cite{Schunck2016Nov,Verriere2020Jul,Regnier2016Mar,Regnier2018Apr,Goutte2004Apr,Regnier2016May,Zdeb2017May,Regnier2019Feb,Regnier2017Sep,Tao2017Aug,Zhao2019May,Zhao2019Jan,Zhao2020Jun,Zhao2021Oct,Zhao2022May,Younes2012Sep,LiZY2022Aug,LiZY2024Jun,RenZX2022AprPRC,LiBo2024Aug} and time-dependent DFT \cite{Simenel2018Nov,Qiang2021Nov,Qiang2021Mar,Pancic2020Sep,Goddard2015Nov,Goddard2016Jan,Scamps2015Jul,Umar2015Aug,Bulgac2019Sep,Scamps2018Dec,RenZX2022Apr,LiBo2024Sep}. The application of TDGCM based on Gogny effective interaction has proven valuable in studying Fm isotopes and capturing the transition from symmetric to asymmetric fission mode \cite{Regnier2019Feb}. In addition, TDGCM method based on covariant DFT has been able to reproduce the triple-peak structure of FFCD of $^{226}$Th \cite{Tao2017Aug,Zhao2019Jan}. Time-dependent DFT calculatioin based on Skyrme functional has revealed that the asymmetric fission mode observed in heavy actinides such as $^{230}$Th, $^{234,236}$U, $^{240}$Pu, $^{246}$Cm, $^{250}$Cf, and $^{258}$Fm, can be attributed to stable octupole deformed shell gap at proton number $Z \approx 56$ for heavy fission fragments \cite{Scamps2018Dec}. 

Despite these advances, a systematically microscopic dynamical study of the Th isotopic chain is lacking. Of particular interest is the role of light fragments (e.g., neutron-rich Kr/Sr isotopes), which exhibit shape coexistence of near spherical and large prolate deformations driven by the deformed proton shell at $Z\approx38$ \cite{Bonatsos2023Aug,Xiang2012Jan,Clement2016Jan,Garrett2022May,Flavigny2017Jun,Gerst2022Feb}. In addition, we note that very recently both theoretical and experimental studies have demonstrated the significant role of $Z=36$ deformed shell on the island of asymmetric fission in the neutron-deficient lead region \cite{Mahata2022Feb,Morfouace2025Apr}, motivating our focused study. Here, we perform a systematic calculation for even-even Th isotopes ($^{222-234}$Th) using the TDGCM based on covariant DFT (CDFT) and analyze the impact of shape coexistence of light fragments on the rapid transition from symmetric to asymmetric fission mode. In order to simulate the excitation environment during fission process, we also consider the finite-temperature effect \cite{Zhao2019Jan}.

\section{THEORETICAL FRAMEWORK}

Nuclear fission is a slow and large-amplitude collective motion and can be described as a collective wave function with some collective degrees of freedom, e.g. axially symmetric quadrupole $\beta_2$ and octupole $\beta_3$ being used here. The low-energy fission dynamics could be simulated by a equation of motion which is derived from the TDGCM in the Gaussian overlap approximation (GOA) \cite{Schunck2016Nov,Verriere2020Jul,Regnier2016May}:
\begin{equation}\begin{aligned}\label{eq:motion}
{\rm i}\hbar\frac{\partial}{\partial t}&g(\beta_2, \beta_3, t) \\
=&\left[-\frac{\hbar^2}{2}\sum_{kl}\frac{\partial}{\partial \beta_k}B_{kl}^{-1}(\beta_2, \beta_3)\frac{\partial}{\partial \beta_l} + V(\beta_2, \beta_3)\right] \\
& \times g(\beta_2, \beta_3, t),
\end{aligned}\end{equation}
where $g(\beta_2, \beta_3, t)$ is a complex wave function of the collective variables $(\beta_2, \beta_3)$ and time $t$, which contains all the information about the dynamics of the system. $V(\beta_2, \beta_3)$ and $B_{kl}(\beta_2, \beta_3)$ are the collective potential and mass tensor, respectively, and they completely determine the dynamics of the fission process in the TDGCM+GOA framework. Following our previous work \cite{Tao2017Aug,LiZY2022Aug,LiZY2024Jun}, the software package {\rm FELIX-2.0} \cite{Regnier2018Apr} is utilized to solve the equation of motion.

The probability current can be defined by the relation
\begin{equation}\begin{aligned}
J_{k}(\beta_2, \beta_3, t) =& \frac{\hbar}{2{\rm i}}\sum_{l=2}^{3}B_{kl}^{-1}(\beta_2, \beta_3)\left[g^{\ast}(\beta_2, \beta_3, t)\frac{\partial}{\partial \beta_l}g(\beta_2, \beta_3, t)\right. \\
& - \left. g(\beta_2, \beta_3, t)\frac{\partial}{\partial \beta_l}g^{\ast}(\beta_2, \beta_3, t)\right]
\end{aligned}\end{equation}
Starting from an initial state of the compound nucleus, the collective current will move to a large deformation region and pass through a so-called scission line that is composed of the hypersurface at which the nucleus splits. At the time t, the measurement of the probability of a given pair of fragments can be calculated when the flux of the probability current runs through the scission hypersurface. For a surface element $\xi $, the sum of the time-integrated flux of the probability $F(\xi, t)$ can be read as \cite{Regnier2018Apr}:
\begin{equation}
    F(\xi, t) = \int_{t=0}^{t} dt \int_{(\beta_2, \beta_3)\in \xi} \bm{J}(\beta_2, \beta_3, t)\, d\bm{S}.
\end{equation}
Each point on the scission line contains the information of ($A_{L}, A_{H}$), which represent the masses of light and heavy fragments, respectively. Hence the yield of fission fragments with mass $A$ can be defined formally as
\begin{equation}
    Y(A) \propto \sum_{\xi \in \mathcal{A} } \lim_{t\rightarrow\infty} F(\xi, t),
\end{equation}
where $\mathcal{A}$ is the set of all element $\xi$ belonging to the scission hypersurface such that the fragments has mass $A$. In this work, we adopt a Gaussian convolution with a constant width $\sigma_{\rm Z} = 2$ to calculate the final charge yields for all Th isotopes.

The dynamics of the equation of motion (\ref{eq:motion}) is determined by the finite-temperature CDFT \cite{Zhao2019Jan}. We firstly construct the entire map of the energy surface in collective space by imposing constraints on the collective coordinates:
\begin{equation}
    \braket{E_{\text{CDFT}}(T)} + \sum_{k=2,3} C_k( \braket{\hat{Q}_k} - q_k)^2, 
\end{equation}
where $\braket{E_{\text{CDFT}}(T)}$ is the total energy of CDFT at temperature $T$ solved based on two-center harmonic oscillator basis \cite{LiZY2024Jun}. $\braket{\hat{Q}_2}$ and $\braket{\hat{Q}_3}$ denote the expectation of mass quadrupole and octupole operators, respectively. $q_k$ is the constrained values of the multipole moments, and $C_{k}$ is the corresponding stiffness constant \cite{Peter1980}. Here, the dimensionless deformation parameters $\beta_k$ derived from $q_k$ \cite{Tao2017Aug} are used. Scission can be described using the Gaussian neck operator $\hat{Q}_N = \exp{[-(z-z_{N})^2/a_{N}^2]}$, where $a_N = 1\, \textrm{fm}$ and $z_N$ is the position of the neck determined by minimizing $\langle \hat{Q}_{N}\rangle$ \cite{Younes2009Nov}. The left and right fragments are defined as parts of the whole nucleus separated at $z_N$.  Following our previous work \cite{Tao2017Aug,LiZY2022Aug,LiZY2024Jun}, the pre-scission domain is defined by $\langle \hat{Q}_N\rangle \geqslant 3$ and consider the frontier of this domain as the scission line. 

\textcolor{red}{The mass tensor in Eq. (\ref{eq:motion}) are calculated in the finite-temperature perturbative cranking approximation \cite{ZhuY2016Aug,Martin2009}:}
\begin{equation}
    B_{kl}(\beta_2, \beta_3) = \hbar^2\left[\mathcal{M}_{(1)}^{-1} \mathcal{M}_{(3)} \mathcal{M}_{(1)}^{-1}\right]_{kl},
\end{equation}
with
\begin{equation}\begin{aligned}
    \mathcal{M}_{(n), kl, T} =& \frac{1}{2}\sum_{i\neq j}\braket{i|\hat{Q}_k|j}\braket{j|\hat{Q}_l|i} \\
    &\left\{\frac{(u_i u_j - v_i v_j)^2}{(E_i - E_j)^n}\left[\tanh\left(\frac{E_i}{2k_B T}\right) - \tanh\left(\frac{E_j}{2k_B T}\right)\right]\right\} \\
    & + \frac{1}{2}\sum_{ij}\braket{i|\hat{Q}_k|j}\braket{j|\hat{Q}_l|i} \\
    &\left\{\frac{(u_i v_j + u_j v_i)^2}{(E_i + E_j)^n}\left[\tanh\left(\frac{E_i}{2k_B T}\right) + \tanh\left(\frac{E_j}{2k_B T}\right)\right]\right\}
\end{aligned}\end{equation}
where $v_i^2$ are the BCS occupation probabilities, and $u_i^2 = 1 - v_i^2$. $E_i$ are the quasiparticle energy. And $k_B$ is the Boltzmann constant.

The thermodynamical potential relevant to study defomation effects is the Helmholtz free energy $F = \braket{E_{\text{CDFT}}(T)} - TS$, evaluated at constant temperature $T$ \cite{Zhao2019Jan,Schunck2015Mar}. Here, $S$ is the entropy of the compound nuclear system.  Then, the collective potential $V(\beta_2, \beta_3)$ is calculated from the free energy by subracting the energy of vibrational zero-point motion $\Delta E_{\rm vib}=\frac{1}{4}{\rm Tr}\left[\mathcal{M}_{(3)}^{-1}\mathcal{M}_{(2)}\right]$ \cite{Girod1979Oct}. Based on the microscopic inputs, we have constructed the equation of motion Eq. (\ref{eq:motion}) to simulate the fission dynamics and calculate the FFCD for Th isotopes. Formalism and numerical details for CDFT and TDGCM can be found in Refs. \cite{LiZY2024Jun} and \cite{Regnier2018Apr}, respectively.

\section{RESULTS AND DISCUSSION}
\begin{figure}[htbp]
    \centering
    \includegraphics[scale=0.88]{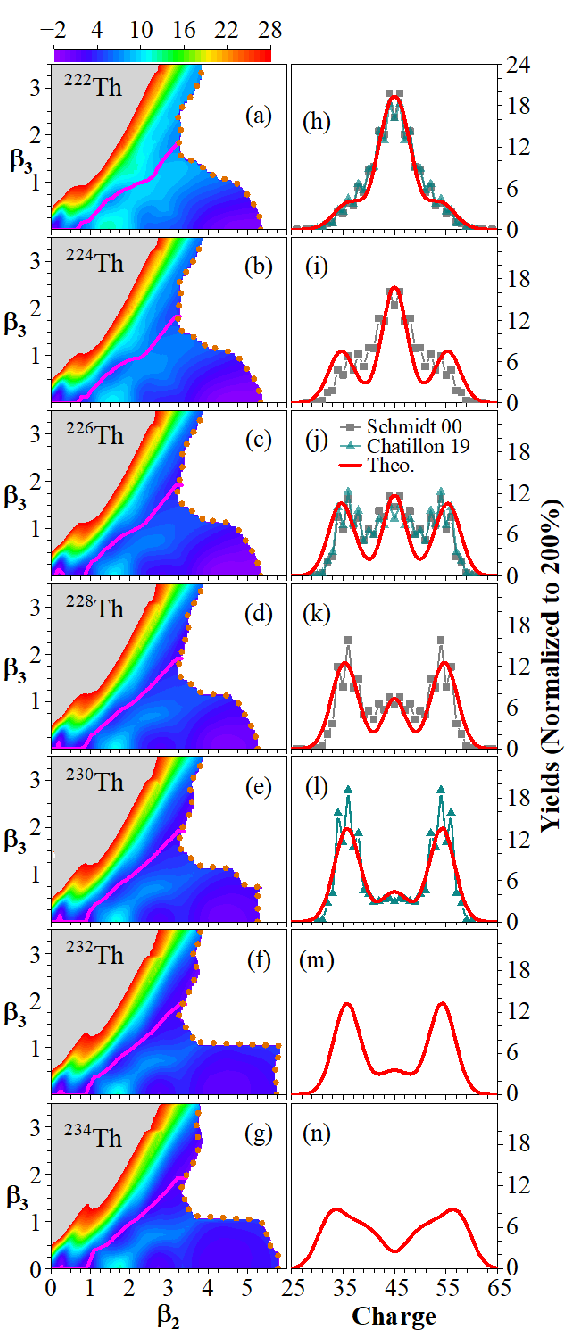}
    \fontsize{11.5pt}{14.4pt}\selectfont
    \caption{(a-g) Collective potential ($F-\Delta E_{\rm vib}$) of even-even $^{222-234}$Th in the $\beta_2$-$\beta_3$ plane calculated by finite-temperature CDFT based on PC-PK1 functional \cite{Zhao2010PRC}. All energies are normalized with respect to energy of the corresponding equilibrium state at $\beta_2\sim0.2$. The optimal fission paths are represented by the magenta solid lines. The orange dotted lines denote the scission lines. (h-n) The FFCDs of $^{222-232}$Th (red solid lines) resulted from TDGCM simulation, in comparison with the experimental data (gray squares from Ref. \cite{Schmidt2000Feb} and cyan triangles from Ref. \cite{Chatillon2019May}). The width of Gaussian convolution used here is $\sigma_z = 2$ for all Th isotopes.}
    \label{fig1:pes-yields}
\end{figure}

\begin{figure*}[htb]
    \centering
    \includegraphics[scale=1.08]{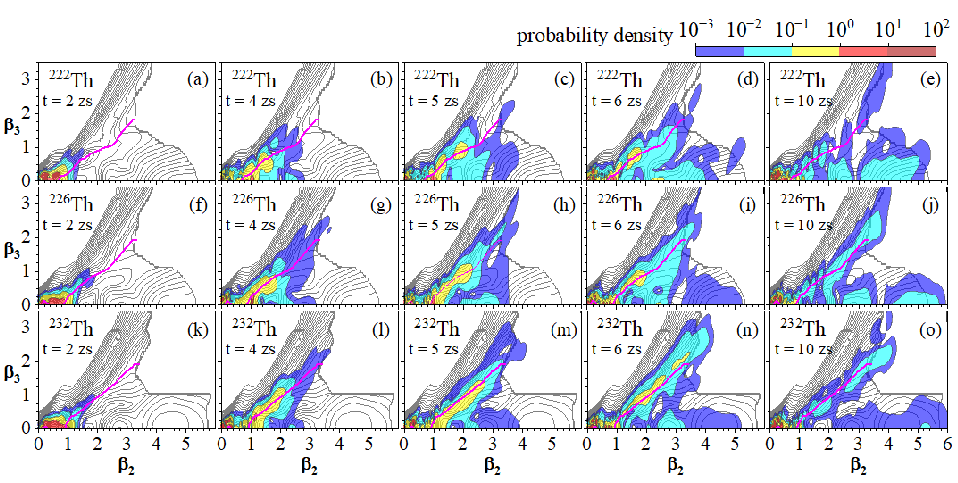}
    \fontsize{11.5pt}{14.4pt}\selectfont
    \caption{Time evolution of the probability density distribution in the $\beta_2-\beta_3$ plane for $^{222, 226, 232}$Th simulated by TDGCM with microscopic inputs determined from finite-temperature CDFT. It is noted that $1 {\rm zs}=10^{-21}$ s.}
    \label{fig2:wvs}
\end{figure*}

Fig. \ref{fig1:pes-yields} (a-g) displays the collective potential of even-even $^{222-234}$Th in the $\beta_2$-$\beta_3$ plane calculated by finite-temperature CDFT. The PC-PK1 functional \cite{Zhao2010PRC} governs the particle-hole channel, while a $\delta$ force with strength parameters $V_n = 360\ {\rm MeV\cdot fm}^3$ and $V_p = 378\ {\rm MeV\cdot fm}^3$ (determined by fitting the empirical pairing gaps of $^{226}$Th \cite{Bender2000Jul}) is employed in the particle-particle channel. The temperatures for $^{222-234}$Th are 0.84, 0.84, 0.83, 0.83, 0.81, 0.79 and 0.77, respectively, corresponding to an excitation energy of 11 MeV in photon-induced fission experiments \cite{Schmidt2000Feb}.

Two distinct fission valleys appear in all collective potential energy surfaces:  An asymmetric fission path beginning at $(\beta_2, \beta_3) \sim (0.8, 0.0)$ and following the magenta solid line to the scission point at $\sim(3.3, 1.9)$; A symmetric fission path along $\beta_3 = 0$. The competition between symmetric and asymmetric fission path in Th isotopes is also observed in other studies \cite{Bernard2020Apr}. Notably, as the neutron number increases, the depth of the symmetric valley remains stable, whereas the asymmetric valley along the optimal path deepens significantly. Consequently, the calculated FFCDs (solid lines in Fig. \ref{fig1:pes-yields} (h-n)) transit sharply from symmetric to asymmetric dominance beyond $^{228}$Th, in excellent agreement with experimental data. 

\begin{figure}[htb]
    \centering
    \includegraphics[scale=0.45]{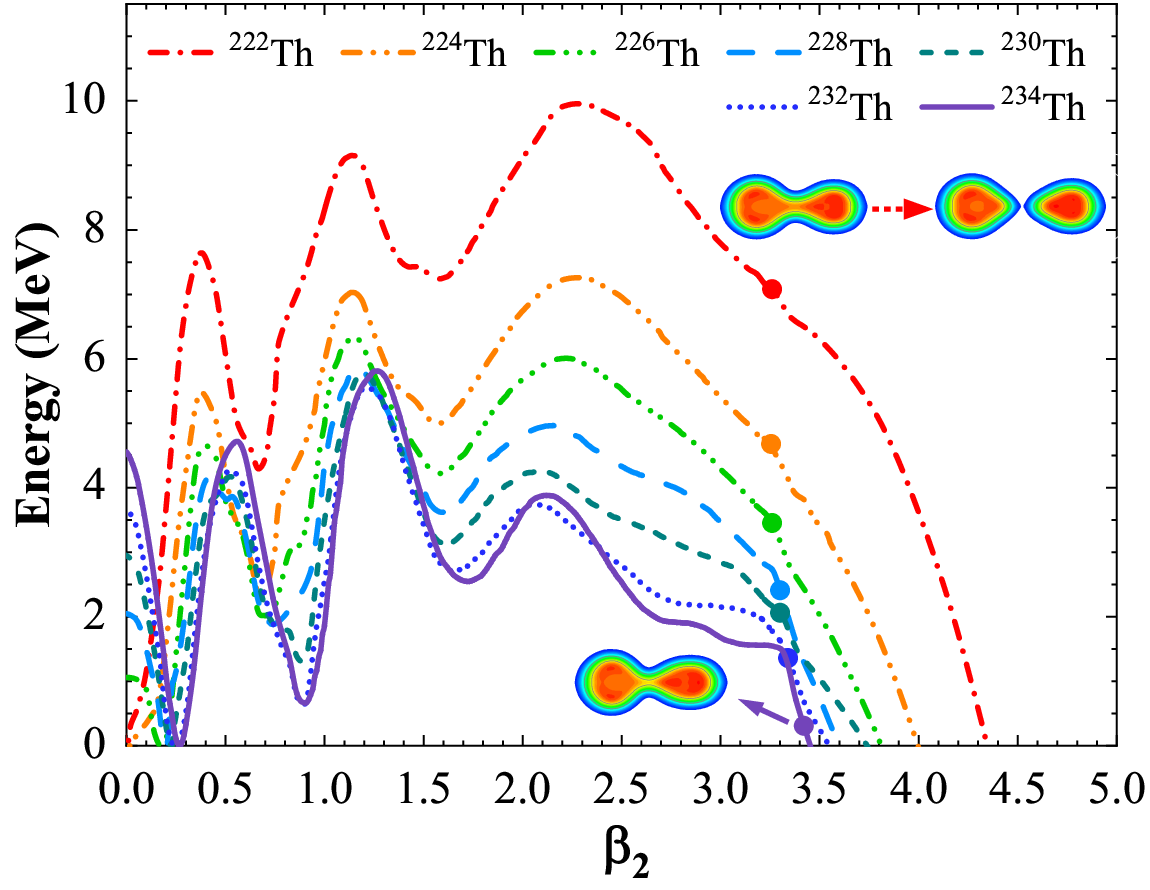}
    \fontsize{11.5pt}{14.4pt}\selectfont
    \caption{Energy curves along the optimal fission paths (magenta solid lines in Fig. \ref{fig1:pes-yields}) for even-even $^{222-234}$Th. The solid circles denote the scission points. Energy curves for post-scission configurations are obtained by performing density constraint calculations with frozen fragments \cite{LiZY2024Jun}. The scission configurations for $^{222, 234}$Th as well as a post-scission configuration for $^{222}$Th are also illustrated. Here we only show some selected configurations since they are all similar.}
    \label{fig3:staticpath}
\end{figure}

To elucidate this rapid transition, we analyze the collective probability density evolution for $^{222,226,232}$Th (Fig. \ref{fig2:wvs}), which exhibit single-symmetric peak, triple peaks, and double-asymmetric peaks of FFCDs, respectively. At $t = 4$ zs, it is evident that all three collective probability waves flow to the asymmetric valley due to the much higher symmetric fission barrier located at $(\beta_2, \beta_3) \sim (1.60, 0.00)$. Over time, the collective probability wave for $^{222}$Th shifts to the lower symmetric valley, yielding a single symmetric peak, while for $^{232}$Th, collective wave persists along the very low asymmetric valley, producing asymmetric-dominated FFCD. In between, for $^{226}$Th, comparable valley depths result in nearly equal flow, generating a triple-peak FFCD.

The above discussion reveals a strong correlation between the rapid fission-mode transition and the evolution of the asymmetric fission valley across the Th isotopic chain. To further clarify this behavior and its underlying mechanism, Fig. \ref{fig3:staticpath} compares the energy curves along the optimal fission paths (magenta solid lines in Fig. \ref{fig1:pes-yields}) for all the Th isotopes, with scission points marked by solid circles. Post-scission configurations are derived from density-constrained calculations with frozen fragments \cite{LiZY2024Jun}. The scission configurations for $^{222, 234}$Th as well as a post-scission configuration for $^{222}$Th are also illustrated. As the neutron number increases, the energy curve drops rapidly until $^{232}$Th then stabilizes for heavier isotopes. Despite an $\sim8$ MeV variation in scission-point heights, all Th isotopes exhibit similar scission configurations: a pear-shaped heavy fragment and a highly deformed light fragment. Given the nearly constant Coulomb energy (fixed proton number and similar scission shapes), the energy drop primarily stems from the variation of neutrons and particularly their partition in the fragments.

\begin{figure}[htb]
    \centering
    \includegraphics[scale=0.72]{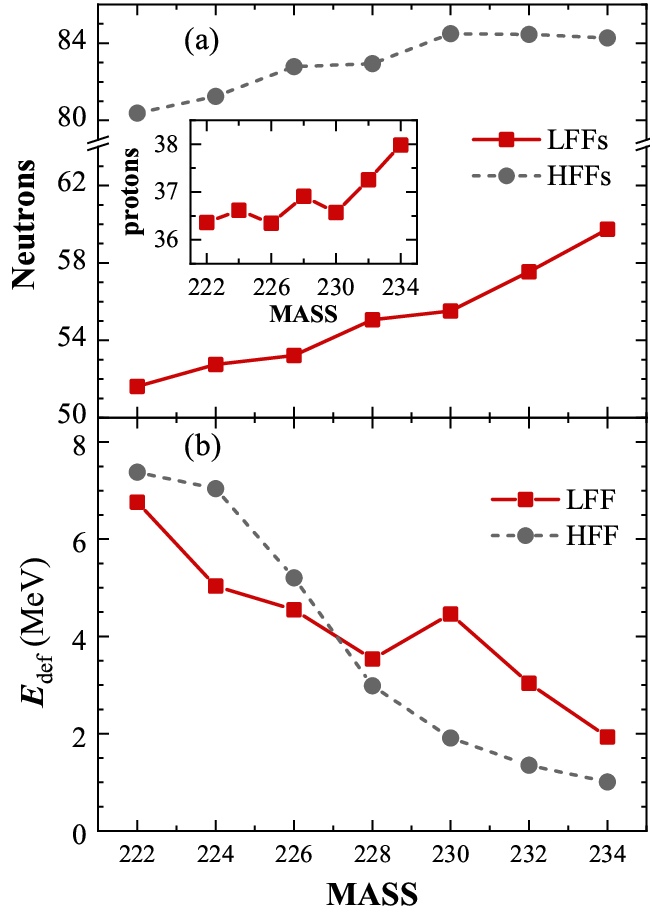}
    \fontsize{11.5pt}{14.4pt}\selectfont
    \caption{Neutron numbers (a) and deformation energies (b) of fission fragments at scission points along Th isotopic chain. Red squres and gray circles denote the results for the light (LFF) and heavy (HFF) fragments, respectively. The inset shows the proton numbers of light fragments.}
    \label{fig4:ffsEdef}
\end{figure}

Therefore, in Fig. \ref{fig4:ffsEdef}, we analyze the properties of the fragments at scission points, specifically the nucleon numbers and deformation energies of fission fragments along Th isotopic chain. To define the deformation energy of fragment more reasonablly, we firstly obtain the post-scission configurations and therefore the separated fragments through density-constrained calculations, as shown in Fig. \ref{fig3:staticpath}. Then, the deformation energy is defined as the energy of the nascent fragment minus its corresponding ground-state energy at the same temperature: $E_{{\rm def}} = E_{{\rm FF}} - E_{{\rm g.s.}}$, as displayed in Fig. \ref{fig4:ffsEdef} (b). Along Th isotopic chain, the neutron numbers for light and heavy fragments increase. In particular, for the light fragment, it rises from $\sim 52$ to $\sim 60$, while the proton number remains in $36\sim38$, consistent with the light asymmetric peak in FFCDs (c.f. Fig. \ref{fig1:pes-yields}). Notably, the deformation energies for both heavy and light fragments exhibit a steep decline: $\sim7.4\to \sim1$ MeV and $\sim6.8\to \sim 2$ MeV for heavy and light fragments, respectively, thereby deepening the asymmetric fission valley. It is well known that the shell effect and octupole deformation effect of heavy fragment play important roles on driving the asymmetric fission of actinides \cite{Schmidt2000Feb,Scamps2018Dec}. However, here we find that not only the heavy fragment but also the light fragment is crucial to drive the rapid transition from symmetric to asymmetric fission mode. A key question arises: What drives the light fragment’s deformation-energy reduction?

\begin{figure}[htb]
    \centering
    {\includegraphics[scale=0.72]{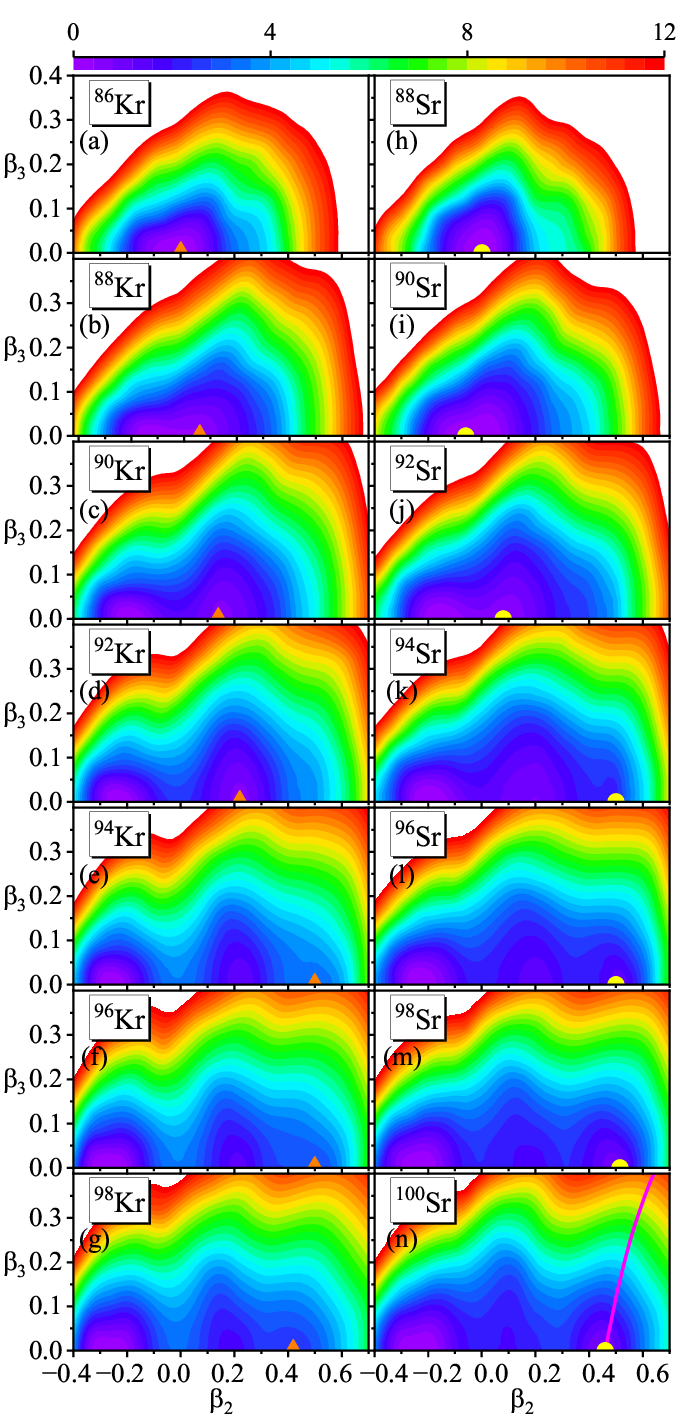}}
    \fontsize{11.5pt}{14.4pt}\selectfont
    \caption{\textcolor{red}{Same as Fig. \ref{fig1:pes-yields} but for even-even $^{86-98}$Kr (a-g) and  $^{88-100}$Sr (h-n) at temperature $0.8~{\rm MeV}$.} All energies are normalized with respect to the corresponding ground state. The orange triangles and yellow circles represent the largest-deformed local minima for Kr and Sr isotopes, respectively.}
    \label{fig5:kr-sr}
\end{figure}

\begin{figure}[htb]
    \centering
    {\includegraphics[scale=0.86]{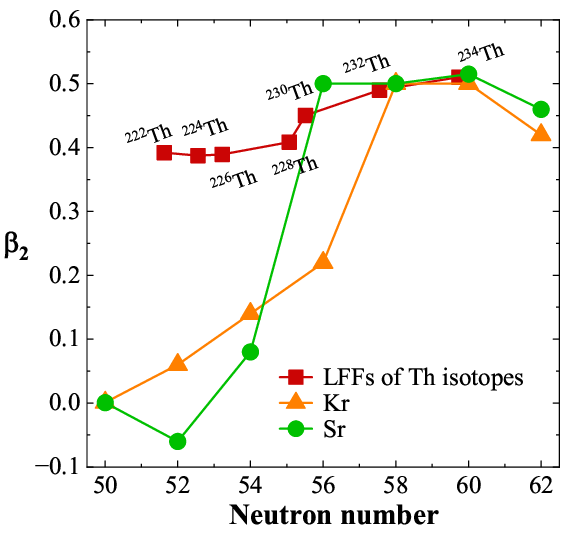}}
    \fontsize{11.5pt}{14.4pt}\selectfont
    \caption{Quadrupole deformations v.s. neutron number for the light fragments at scission points for Th isotopes (red squres), in comparison with those of largest-deformed minima in Kr (orange triangle) and Sr (green circles) isotopes. Note that the neutron number for light fragment is not exactly integer since it is calculated from its intrinsic neutron density distribution. This can be solved by incorporating particle number projection techniques \cite{Verriere2021PRC}.}
    \label{fig6:beta2-min}
\end{figure}

To address this question, Fig. \ref{fig5:kr-sr} shows the collective potential in the $\beta_2$-$\beta_3$ plane for Kr ($Z=36$) and Sr ($Z=38$) isotopes --- the dominant species in the light asymmetric FFCD peak --- calculated with temperature $T=0.8~{\rm MeV}$, which is close to those of the compound nuclei. Spherical shape is observed for $N=50$ isotones due to the shell effect. With increasing neutron number ($N \geq 54$), collective potential energy surfaces soften along quadrupole ($\beta_2$) and octupole ($\beta_3$) deformations, developing multiple minima. Notably, well-deformed local minima ($\beta_2 > 0.4$) or even global minima emerge in $^{96,98}$Kr and $^{94-100}$Sr, which is supported by the measured charge radii and low-lying spectra \cite{Mach1989Jan,Mach1991Sep,Buchinger1990Jun,Angeli2004May,RzacaUrban2009Feb,Schussler1980May,Flavigny2017Jun,Gerst2022Feb}. Furthermore, we note that the quadrupole deformations of these minima are consistent with those of the light fragments at scission points, shown in Fig. \ref{fig6:beta2-min}. This match allows Kr/Sr nuclei to accommodate light-fragment stretching efficiently, reducing deformation energy in asymmetric fission ($^{230}$Th and heavier isotopes in Fig. \ref{fig4:ffsEdef}). In contrast, lighter Th isotopes exhibit larger deformation mismatches between their light fragments and nearby Kr/Sr isotopes, raising deformation energy and asymmetric-fission valley heights (Figs. \ref{fig4:ffsEdef} and \ref{fig3:staticpath}).

\begin{figure}[htp]
    \centering
    {\includegraphics[scale=0.86]{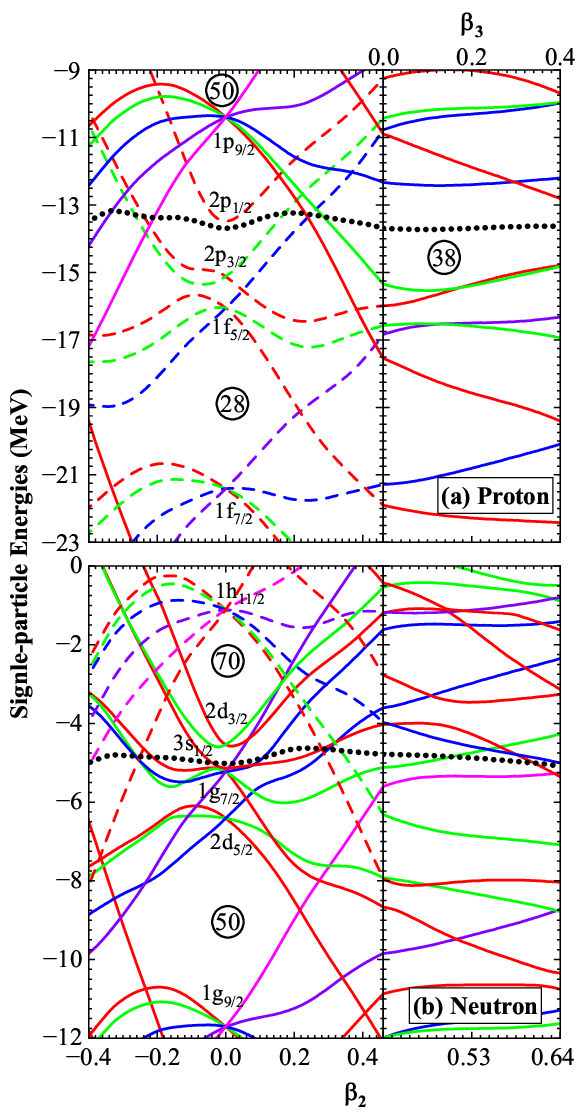}}
    \fontsize{11.5pt}{14.4pt}\selectfont
    \caption{Single-particle energies as a function of the quadrupole ($\beta_2$) and octupole ($\beta_3$) deformations for the protons (a) and neutrons (b) in $^{100}$Sr. Left parts show the results for states with $\beta_3 = 0$, while right parts for those following the magenta solid curve in Fig. \ref{fig5:kr-sr} (n). The dotted curves denote the Fermi levels.}
    \label{fig7:Srlevels}
\end{figure}

Finally, in Fig. \ref{fig7:Srlevels}, we plot the single-particle energies as a function of the quadrupole ($\beta_2$) and octupole ($\beta_3$) deformations for the protons and neutrons in $^{100}$Sr as an example to illustrate the origin of shape coexistence in this mass region. Left panels show the results for states with $\beta_3 = 0$, while right ones for those following the magenta solid curve in Fig. \ref{fig5:kr-sr} (n). The dotted curves denote the Fermi levels. We find that both proton and neutron shells around $\beta_2=-0.2$ give rise to the oblate minimum. Notably, a deformed proton $Z=38$ shell emerges around $\beta_2 \approx 0.46$ and lasts along the octupole deformation till $\beta_3\approx0.3$, which causes the large deformed prolate minimum and softness against octupole deformation in this mass region. 

\section{SUMMARY AND OUTLOOK}

In summary, we have systematically investigated the evolution of fission modes along the Th isotopic chain using a microscopic framework combining the time-dependent generator coordinate method (TDGCM) and finite-temperature covariant density functional theory with the relativistic PC-PK1 functional. Our theoretical fission fragment charge distributions exhibit excellent agreement with experimental data, except for the odd-even staggering effect. Both calculations and data reveal a rapid transition from symmetric to asymmetric fission as the mass number increases from $A = 222$ to 234.  By analyzing the collective potential energy surfaces and time evolution of collective probability density distributions, we demonstrate that this rapid transition of fission mode is strongly correlated with the rapidly deepening asymmetric fission valley, $\sim8$ MeV for the scission configuration from $^{222}$Th to $^{234}$Th. This is driven by the reduction of deformation energies of both the heavy and light fragments: $\sim7.4\to \sim1\ {\rm MeV}$ and $\sim6.8\to \sim 2\ {\rm MeV}$ for the heavy and light fragments, respectively. Further analysis attributes the decrease of light-fragment deformation energies to the onset of a coexisting large-deformed minimum in neutron-rich Kr and Sr isotopes (dominated isotopes for light asymmetric peak), which arises from a deformed proton $Z=38$ shell closure near $\beta_2\approx0.46$.  Notably, we identify, for the first time, the pivotal role of the light fragment and its shape coexistence structure on the fission mode transition in Th isotopes in a fully microscopic framework. 

In the future, we may incorporate particle number projection techniques \cite{Verriere2021PRC} within the CDFT framework to improve the description of odd-even effects in fission fragment charge distributions. In addition, we have implemented Fourier shape parametrization in constrained CDFT very recently \cite{LiZY2025PLB}, which offers significant advantages for modeling highly elongated configurations. This method generates smooth, minimally correlated 3D potential energy surface (PES), enabling high-precision dynamical simulations of fission. Therefore, it will be interesting to perform a systematic dynamical calculation for Th and also other isotopes based on 2D or even 3D Fourier shape space.

\section*{ACKNOWLEDGMENTS}
This work was partly supported by the National Natural Science Foundation of China (Grants No. 12375126), the Fok Ying-Tong
Education Foundation, and the Fundamental Research Funds for the Central Universities.



\bibliographystyle{apsrev4-1}
\bibliography{apssamp}





\end{document}